\documentclass[twocolumn,aps,prl]{revtex4}
\usepackage{graphicx}

\begin{document}
\bibliographystyle{prsty}

\title{Non-Locality and Strong Coupling in the Heavy Fermion Superconductor CeCoIn$_{5}$:
A Penetration Depth Study}
\author{Elbert E. M. Chia}
\author{D. J. Van Harlingen}
\author{M. B. Salamon}
\affiliation{Department of Physics, University of Illinois at
Urbana-Champaign, 1110 W. Green St., Urbana IL 61801}
\author{Brian D. Yanoff}
\affiliation{General Electric, Schenectady, NY}
\author{I. Bonalde}
\affiliation{Centro de F\'{i}sica, Instituto Venezolano de
Investigaciones Cient\'{i}ficas, Apartado 21827, Caracas 1020-A,
Venezuela}
\author{J. L. Sarrao}
\affiliation{Los Alamos National Laboratory, MST-10, Los Alamos,
NM 87545}
\date{\today}

\begin{abstract}
We report measurements of the magnetic penetration depth $\lambda$
in single crystals of CeCoIn$_{5}$ down to $\sim$0.14~K using a
tunnel-diode based, self-inductive technique at 28 MHz. While the
in-plane penetration depth tends to follow a power law,
$\lambda_{//} \sim {\it T}^{3/2}$, the data are better described
as a crossover between linear ({\it T} $\gg $ ${\it T}^\ast $) and
quadratic ({\it T} $\ll {\it T}^\ast $) behavior, with ${\it
T}^\ast $ the crossover temperature in the strong-coupling limit.
The {\it c}-axis penetration depth $\lambda_{\perp} $ is linear in
{\it T}, providing evidence that CeCoIn$_{5}$ is a {\it d}-wave
superconductor with line nodes along the {\it c}-axis. The
different temperature dependences of $\lambda_{//} $ and
$\lambda_{\perp} $ rule out impurity effects as the source of
${\it T}^{\ast} $.
\end{abstract}

\maketitle
The compounds CeMIn$_{5}$ (M = Co, Ir, Rh) have recently
been added to the heavy-fermion family, and have attracted much
interest due to their similarity with the cuprates: quasi-2D
structure and proximity to magnetic order \cite{Petrovic2001}.
CeCoIn$_{5}$, in particular, is a good candidate for study: its
superconductivity is not sensitive to small changes in unit-cell
volume or composition, unlike CeCu$_{2}$Si$_{2}$, and it has the
highest {\it T}$_{c}$ ($\sim $2.3~K) among the heavy-fermion
superconductors. CeCoIn$_{5}$ has tetragonal HoCoGa$_{5}$ crystal
structure, consisting of alternating layers of CeIn$_{3}$ and
'CoIn$_{2}$' \cite{Petrovic2001}. De Haas-van Alphen (dHvA) data
revealed that the Fermi surface (FS) is quasi-2D, with an open 2D
undulating cylinder extending along the [001] direction, as well
as the large effective masses of electrons on this FS
\cite{Hall2001}.

Recently, there has been mounting evidence for unconventional
superconductivity in CeMIn$_{5}$. Specific heat data reveal a {\it T%
}$^{2}$ term at low temperature, consistent with the presence of
line nodes in the superconducting energy gap
\cite{Movshovich2001}. Thermal conductivity measurements with
in-plane applied field show four-fold symmetry, consistent with
nodes along the ($\pm \pi $, $\pm \pi $) positions
\cite{Izawa2001}. NQR measurements show that there is no
Hebel-Slichter peak just below {\it T}$_{c}$ \cite{Kohori2001}.
Below ${\it T_{c}}$ the spin susceptibility is suppressed,
indicating singlet pairing \cite{Kohori2001,Curro2001}. However,
there are some ambiguities in some of the measurements. Thermal
conductivity data yield a {\it T}$^{3.37}$ low-temperature
behavior, that the authors claim is close to T$^{3}$ behavior
predicted for unconventional superconductors with line nodes in
the clean limit \cite{Movshovich2001}. NQR measurements did not
show the ${\it T^{3}}$ low-temperature behavior of 1/${\it T_{1}}$
that is expected for a line node gap; instead 1/${\it T_{1}}$
saturates below 0.3~K \cite{Kohori2001}. Microwave measurements
down to $\sim $0.2~K showed a non-exponential behavior, and the
authors claimed that $\lambda $({\it T}) $\sim ${\it T} below
0.8~K \cite{Ormeno2002}, though the data clearly show some
curvature in that temperature range. Further, the field was
applied along the {\it ab}-plane, so the shielding currents have
both in-plane and inter-plane components. In this paper, we
present high-precision measurements of in-plane $\lambda _{//}$
and inter-plane $\lambda _{\perp }$ penetration depths of
CeCoIn$_{5}$ at temperatures down to 0.14~K. We find that
$\lambda_{//} $ is best treated as a crossover from $\sim ${\it T}
to $\sim ${\it T}$^{2}$ at a temperature ${\it T}^{\ast} $.
Combined with the result that $\lambda_{\perp} \propto {\it T} $,
this gives strong evidence for non-local behavior in a {\it
d}-wave superconductor as predicted by Kostzin and Leggett
\cite{Kosztin1997}.

Details of sample growth and characterization are described in
Refs. \cite{Petrovic2001,Petrovic2001b}. Measurements were
performed utilizing a 28 MHz tunnel diode oscillator
\cite{Bonalde2000} with a noise level of 1 part in 10$^{9}$ and
low drift. The magnitude of the ac field was estimated to be less
than 5 mOe. The cryostat was surrounded by a bilayer Mumetal
shield that reduced the dc field to less than 1 mOe. The sample
was aligned inside the probing coil in two directions: (1) {\it
ab} plane perpendicular to the rf field, measuring the in-plane
penetration depth $\lambda _{//}$ (screening currents in the {\it
ab} plane); or (2) with the rf field parallel to the plane, giving
a combination of $\lambda _{//}$ and $\lambda _{\perp}$. The
sample was mounted, using a small amount of GE varnish, on a rod
made of nine thin 99.999\% Ag wires embedded in Stycast 1266
epoxy. The other end of the rod was thermally connected to the
mixing chamber of an Oxford Kelvinox 25 dilution refrigerator. The
sample temperature is monitored using a calibrated RuO$_{2}$
resistor at low temperatures ({\it T}$_{base}$ - 1.8~K), and a
calibrated Cernox thermometer at higher temperatures (1.3~K -
2.5~K). We report data only for ${\it T} \geq $ 0.14~K. The value
of {\it T}$_{c}$ was determined from magnetization measurements to
be 2.3~K, identical to the previously reported value [3].

The deviation $\Delta \lambda $({\it T}) = $\lambda $({\it T}) -- $%
\lambda $(0.14~K) is proportional to the change in resonant
frequency $\Delta ${\it f}({\it T}), with the proportionality
factor {\it G} dependent on sample and coil geometries. For a
square sample of side 2{\it w}, thickness 2{\it d},
demagnetization factor {\it N},
and volume {\it V}, {\it G} is known to vary as {\it G} $%
\propto $ {\it R}$_{3D}$(1-{\it N})/{\it V}, where {\it R}$_{3D}$
= {\it w}/[2(1+(1+2{\it d}/{\it w})$^{2}$)$\arctan$({\it w}/2{\it d})-2{\it d}/{\it w%
}] is the effective sample dimension \cite{Prozorov2000}. For our
sample 2{\it w}~$\approx $~0.73~mm and 2{\it d}~$\approx
$~0.09~mm. We determine {\it G} from a single-crystal sample of
pure Al by fitting the Al data to extreme non-local expressions
and then adjusting for relative sample dimensions. Testing this
approach on a single crystal of Pb, we found good agreement with
conventional BCS expressions.

\begin{figure}
\centering
\includegraphics[width=8cm,clip]{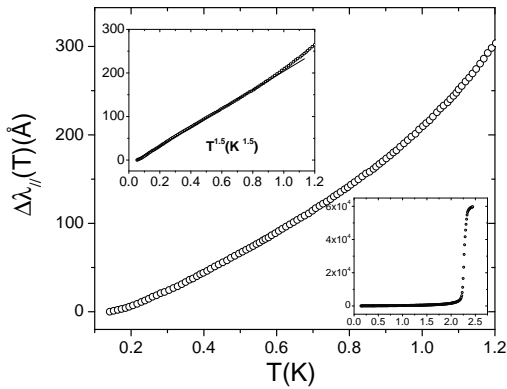}
\caption{Low-temperature dependence of the in-plane penetration
depth $\Delta \lambda _{//}$(%
{\it T}). Lower inset shows $\Delta \lambda _{//}$(%
{\it T}) over the full temperature range. Upper inset shows $\Delta \lambda _{//}$(%
{\it T}) vs {\it T}$^{1.5}$ in the temperature range
(0.14-1.13)~K. The solid line is a guide to the eye.}
\label{fig:lambdaab}
\end{figure}

Fig.~\ref{fig:lambdaab} shows $\Delta \lambda_{//} $({\it T}) as a
function of temperature. We see that $\Delta \lambda_{//} $({\it
T}) varies strongly at low temperatures, inconsistent with the
exponential behavior expected for isotropic {\it s}-wave
superconductors. On the other hand, the variation is not linear,
but has an obvious upward curvature, unlike the low-temperature
behavior expected for pure {\it d}-wave superconductors. A fit of
the low temperature data to a variable power law, $\Delta \lambda
_{//}$(T) = {\it a} + {\it bT}$^{n}$ yields {\it n} = 1.43 $\pm $
0.01 for sample 1 and 1.57 $\pm $ 0.01 for sample 2. The upper
inset of Fig.~\ref{fig:lambdaab} shows this approximate {\it
T}$^{3/2}$ behavior for sample 1. Kosztin {\it et. al.}
\cite{Kosztin1998,QChen1998,QChen9908362} have proposed a theory
that gives a ${\it T}^{3/2} $ term from the gradual evolution of
the
pseudogap above ${\it T}_{c} $ to the superconducting gap below {\it T}$%
_{c}$. While resistivity measurements suggest the
possibility of a pseudogap in CeCoIn$_{5}$ \cite{Sidorov2002},
which renders this interpretation feasible, a decrease in
Knight shift was observed only starting at ${\it T_{c}}$
\cite{Curro2001}. We take the latter to rule out a pseudogap mechanism.

Before considering novel excitation processes, we note the
important distinction between $\Delta \lambda $({\it T}), which is
directly measured, and the superfluid density [$\rho $({\it T}) =
$\lambda ^{2}$(0)/$\lambda ^{2}$({\it T})] which can be inferred
only with the knowledge of $\lambda $(0) \cite{Carrington1999}. In
the {\it d}-wave model, even if $\rho $ varies strictly with {\it
T}, i.e. $\rho $=1-$\alpha ${\it T}/{\it T}$_{c}$, the penetration
depth is non-linear: $\lambda $({\it
T}) = $\lambda $(0)[1+1/2 ($\alpha ${\it T}/{\it T}$_{c}$) + 3/8 ($\alpha $%
{\it T}/{\it T}$_{c}$)$^{2}$+...]. Hence there is always a
quadratic component to $\lambda $ whose strength depends on
$\alpha $, which in the {\it d}-wave model, is inversely
proportional to ${d\Delta (\theta )/d\theta |_{node}}$, the
angular slope of the energy gap at the nodes \cite{Xu1995}. If
$\rho $({\it T}) is linear in {\it T}, there is no need to invoke
another mechanism.

To extract the in-plane superfluid density from our data, we need
to know $\lambda _{//}$(0). For a quasi-2D superconductor with a
cylindrical Fermi surface and the material parameters in
Ref.~\onlinecite{Movshovich2001} \cite{Barash2000}, we obtain
$\lambda _{//}$(0) = 2600~\AA, considerably larger than the
experimentally obtained value of 1900~\AA\ \cite{Ormeno2002}. This
along with a large heat-capacity jump at ${\it T}_{c}$ leads us to
consider strong-coupling corrections as listed below
\cite{Orlando1979, Kresin1975}:

\begin{equation}
\eta _{Cv}(\omega _{0})=1+1.8(\frac{\pi T_{c}}{\omega _{0}})^{2}(\ln (\frac{%
\omega _{0}}{T_{c}})+0.5); \label{eqn:strongcouplingCv}
\end{equation}%

\begin{equation}
\eta _{\Delta }(\omega _{0})=1+5.3(\frac{T_{c}}{\omega _{0}})^{2}\ln (\frac{%
\omega _{0}}{T_{c}}); \label{eqn:strongcouplingDelta}
\end{equation}%

\begin{equation}
\eta _{\lambda }(\omega _{0})=\frac{\sqrt{1+(\frac{\pi T_{c}}{\omega _{0}}%
)^{2}(0.6\ln (\frac{\omega _{0}}{T_{c}})-0.26)}}{1+(\frac{\pi
T_{c}}{\omega _{0}})^{2}(1.1\ln (\frac{\omega _{0}}{T_{c}})+0.14)};
\label{eqn:strongcouplinglambda}
\end{equation} each $\eta $ represents the correction to the corresponding BCS
value. If we take the experimental value of $\Delta $C/$\gamma
${\it T$_{c}$} = 4.5 \cite{Petrovic2001}, then
Eq.~\ref{eqn:strongcouplingCv} gives the characteristic
(equivalent Einstein) frequency $\omega_{0}$ = 9.1~K and
$\lambda_{//}^{sc} $(0) = 1500~\AA. However, Petrovic {\it et.
al.} \cite{Petrovic2001} argued that since {\it C/T} increases
with decreasing temperature, the specific heat coefficient $\gamma
$ is temperature-dependent below ${\it T_{c}}$. This effect calls
into question simple estimates of strong-coupling corrections for
CeCoIn$_{5}$. A better estimate is to use $\Delta C$/$\Delta S$,
where $\Delta S$ is the measured change in entropy of the sample
from {\it T} = 0 to ${\it T_{c}}$. Ref.~\onlinecite{Petrovic2001}
then gives $\Delta C$/$\Delta S$ = 2.5, so that $\omega_{0} $ =
17.9~K, resulting in $\Delta_{0}^{sc} $ = 2.1{\it k$_{B}$T$_{c}$}
and $\lambda_{//}^{sc} $(0) = 2000 \AA. On the other hand, the
larger $\Delta ${\it C} of Ref.~\onlinecite{Ikeda2001} yields
$\Delta C$/$\Delta S$ = 2.8 and $\omega_{0} $ = 15.4~K, leading to
$\Delta_{0}^{sc} $ = 2.2{\it k$_{B}$T$_{c}$} and
$\lambda_{//}^{sc} $(0) = 1900 \AA. These values of
$\lambda_{//}^{sc} $(0) are close to that obtained by Ormeno {\it
et. al.} \cite{Ormeno2002}.

\begin{figure}
\centering
\includegraphics[width=8cm,clip]{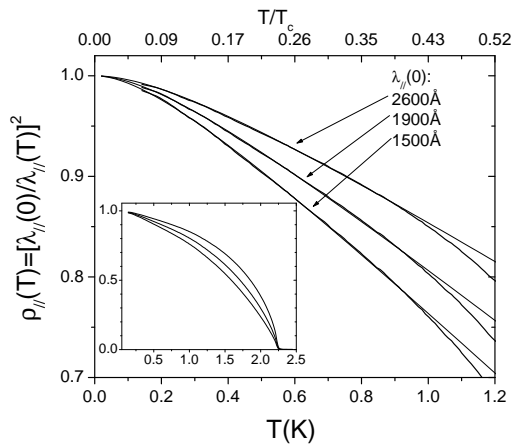}
\caption{ Low-temperature in-plane superfluid density $\rho_{//}
$(T) =
[$\lambda_{//}^{2}$(0)/$\lambda_{//}^{2}$(T)] calculated from $\Delta \lambda _{//}$(%
{\it T}) data in Fig. 1 (thick lines). The thin lines correspond
to fits to data using Eq.~\ref{eqn:hirschfeldns}, using three
values of $\lambda _{//}$(0). Inset shows $\rho_{//}$(T) over the
full temperature range.} \label{fig:hirschfeld}
\end{figure}

Although we will argue that non-local effects are important, we
will refer to $(\lambda_{//}(0)/\lambda_{//}(T))^{2} $ as the
``superfluid density.'' Fig.~\ref{fig:hirschfeld} shows the
calculated behavior of that quantity using the three values of
$\lambda_{//} $(0) obtained above. We follow the procedure in
Ref.~\onlinecite{Carrington1999} to compute the experimental
superfluid density, using the {\it T}$^{3/2}$ fit to estimate the small difference between $%
\lambda _{//}$(0) and $\lambda _{//}$(0.14~K). In each case, $\rho
$({\it T}) is clearly not linear in {\it T}.

Non-linearity in $\rho $({\it T}) can arise from a crossover from
an intermediate-temperature (pure) linear-{\it T} behavior to, for
example, low-temperature (impurity-dominated) quadratic behavior
as pointed out by Hirschfeld and Goldenfeld \cite{Hirschfeld1993}.
They interpolated between these two regions using

\begin{equation}
\lambda  = \lambda _{0} + {\it bT}^{2}/({\it T}^{\ast }+{\it T}),
\label{eqn:hirschfeldlambda}
\end{equation} where {\it T}$^{\ast }$ is the crossover temperature.
In terms of superfluid density, one obtains \cite{Carrington1999}

\begin{equation}
\Delta \rho _{//}(T)=\frac{\alpha T^{2}/T_{c}}{T^{\ast }+T},
\label{eqn:hirschfeldns}
\end{equation}where ${\it T}^{\ast} $ depends on impurity concentration.

A much more provocative source of the crossover of
Eq.~\ref{eqn:hirschfeldns} was suggested by Kosztin and Leggett
[KL] \cite{Kosztin1997}, who showed that for {\it d}-wave
superconductors, nonlocal effects change the linear behavior to
quadratic below a crossover temperature {\it T}$_{nonlocal}^{\ast
}$ = $\Delta_{0} \xi _{//}$(0)/$\lambda_{//}$(0).

The solid lines in Fig.~\ref{fig:hirschfeld} are fits to
Eq.~\ref{eqn:hirschfeldns} and are very good for all three values
of $\lambda _{//}$(0). The value of $\alpha $ varies from $\sim
$0.5 to 0.7, the smallest value of $\alpha $ belonging to the
largest value of $\lambda _{//}$(0). The value of $\alpha $
obtained is similar to that found for YBa$_{2}$Cu$_{3}$O$_{6.95}$
($\alpha \sim $~0.6) \cite{Hardy1993,Carrington1999b}, but smaller
than that of
Tl$_{2}$Ba$_{2}$CuO$_{6+\delta }$ ($\alpha \sim $~1.0) \cite{Broun1997} and K-(ET)$%
_{2}$Cu[N(CN)$_{2}$]Br ($\alpha \sim $~1.2) \cite{Carrington1999}.
The value of {\it T}$^{\ast }$ varies less across the three
$\lambda _{//}$(0) values, from 0.32~K to 0.42~K. These values of
{\it T}$^{\ast }$/{\it T}$_{c}$ ($\sim $0.14 - 0.18) differ from
the cuprates \cite{Bonn1994,Hardy1993,Broun1997} and the organic
superconductor K-(ET)$_{2}$Cu[N(CN)$_2$]Br ($\sim $0.05), where
impurity scattering is presumed to be the source. Further,
Ref.~\onlinecite{Movshovich2001} puts an upper limit of 20~ppm on
the impurity concentration. In the dirty {\it d}-wave model
\cite{Hirschfeld1993}, this gives the unitary-limit scattering
rate $\Gamma \sim $~1.5 $\times $ 10$^{8}$~s$^{-1}$, which yields
an upper limit for {\it T}$^{\ast } \sim $~65~mK. This is about 5
times smaller than the experimentally obtained values above,
suggesting that the sample is too clean for the dirty {\it d}-wave
model to be applicable.

\begin{figure}
\centering
\includegraphics[width=8cm,clip]{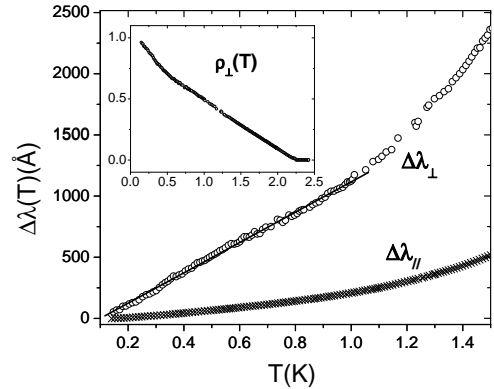}
\caption{Low-temperature dependence of inter-plane (open circles) penetration depth $\Delta \lambda _{\perp}$(%
{\it T}), after subtracting the in-plane component. In-plane $\Delta \lambda _{//}$(%
{\it T}) (crosses) data is also shown for comparison. Solid line
is a linear fit from 0.14~K to 1~K. Inset shows inter-plane
superfluid density $\rho_{\perp}$({\it T}) for the whole
temperature range.} \label{fig:nsperp}
\end{figure}

Having ruled out impurity scattering, we turn to nonlocal
electrodynamics as the source of the crossover in $\rho_{//}(T) $.
For a {\it d}-wave superconductor {\it with line nodes along the
{\it c}-axis}, nonlocality is expected to be relevant only when
the applied magnetic field is oriented parallel to the {\it
c}-axis, while the effect of impurities should not depend on the
orientation of the field. As KL noted, if {\it T}$^{\ast }$ is
noticeably smaller for ${\it H}\perp{\it c}$ than for ${\it
H}//{\it c}$ we may conclude that the observed effect is due
mainly to nonlocal electrodynamics and not to impurities. For
${\it H}\perp{\it c}$, screening currents flow both parallel and
perpendicular to the {\it c}-axis, mixing $\lambda_{//}$ and
$\lambda_{\perp}$ with the frequency shift given by
$\frac{\Delta f_{\perp }}{f_{0}}=\frac{V}{2V_{0}}(\frac{\lambda _{//}}{d}+%
\frac{\lambda _{\perp}}{w})$ \cite{Prozorov2000}, where ${\it
V}_{0}$ is the effective coil volume and ${\it f}_{0}$ the
resonant frequency with the sample absent. In order to extract
$\lambda_{\perp} $ we subtract out the $\lambda_{//} $ component
from $\Delta f_{\perp}$. Fig.~\ref{fig:nsperp} shows the
inter-plane penetration depth $\lambda_{\perp} $ of CeCoIn$_{5}$
down to 0.14~K. It is clearly linear in {\it T} from 0.14~K to
1~K. To obtain the superfluid density, we estimate
$\lambda_{\perp} $(0) from the {\it H}$_{c2}$ anisotropy of $\sim
$2.3 \cite{Ikeda2001}, and the fact that $\lambda $(0) $\propto
\sqrt{{\it H_{c2}}(0)} $ \cite{Gross1986}, obtaining $\lambda
_{\perp }$(0) $\sim $ 2700~\AA. This is close to the value of
$\sim $2800~\AA\ obtained from microwave measurements in the
planar geometry \cite{Ozcan2002}. If we fit $\lambda_{\perp}
$({\it T}) to Eq.~\ref{eqn:hirschfeldlambda}, we find {\it
T}$_{\perp }^{\ast } \lesssim $ 0.15~K, significantly smaller than
0.32~K obtained for the in-plane case. This satisfies the
Kosztin-Leggett test and indicates that the superfluid response of
CeCoIn$_5$ is governed by nonlocal electrodynamics. This is also
strong evidence that {\it CeCoIn$_{5}$ is a {\it d}-wave
superconductor with line nodes along the {\it c}-axis}.
Sr$_{2}$RuO$_{4}$ failed this test \cite{Bonalde2000} because its
line nodes are horizontal instead of vertical. Kusunose and
Sigrist argued that horizontal line nodes give power-law behaviors
with less angular dependence for any inplane direction of the
screening currents, and hence applied field \cite{Kusunose2002}. A
calculation of $\rho_{\perp} $ is shown in the inset of
Fig.~\ref{fig:nsperp}: the upturn below 0.5~K is an artifact of
the choice of $\lambda_{c}$(0). A larger value of $\lambda_{c}$(0)
would remove this feature, but there is no justification for doing
so.

As a final test of the non-local scenario, we estimate {\it
T}$_{//}^{\ast }$ using strong-coupling parameters. From the
measured {\it H$_{c2}$}(0)[001] value of 49.5~kOe, the coherence
length $\xi_{//}$(0) is calculated to be 82 \AA\ \cite{Ikeda2001}.
Together with the earlier-derived values of $\Delta_{0}^{sc}
$~=~2.2{\it k$_{B}$T$_{c}$} and $\lambda
_{//}^{sc}$(0)~=~1900~\AA, we find the strong-coupling nonlocal
crossover temperature {\it T}$_{nonlocal}^{\ast }$ =
$\Delta_{0}^{sc}~\xi _{//}$(0)/$\lambda _{//}^{sc}$(0) = 0.22~K.
Using a weak-coupling {\it d}-wave $\Delta $(0) = 2.14{\it
k$_{B}$T$_{c}$}, we find {\it T}$_{nonlocal}^{\ast }$ = 0.26~K. We
regard either value to be satisfactorily close to the
experimental value of 0.32~K. Note that the value of $\xi_{//}$(0)
is different from the calculated BCS value of 58 \AA\
\cite{Movshovich2001} or the strong-coupling corrected value of
$\sim $ 50 \AA\ \cite{Orlando1979}. This is not surprising since
the BCS expressions \cite{Orlando1979} assume a spherical FS,
while LDA band structure reveals a very complicated FS with
contributions from three different bands \cite{Wills}.

In conclusion, we report measurements of the magnetic penetration
depth $\lambda$ in single crystals of CeCoIn$_{5}$ down to
$\sim$0.14~K using a tunnel-diode based, self-inductive technique
at 28 MHz. The in-plane penetration depth ($\lambda$$_{//}$)
exhibits a crossover between linear (at high T) and quadratic (low
T) behavior with a crossover temperature {\it T}$_{nonlocal}^{\ast
} \approx $ 0.32~K. Such behavior can arise in a superconductor
with nodes in the gap either in a dirty {\it d}-wave model or from
nonlocal electrodynamics. The linear low-temperature dependence of
the {\it c}-axis penetration depth $\lambda_{\perp}$ strongly
favors the nonlocal model with line nodes parallel to the {\it
c}-axis. We also demonstrate that strong-coupling corrections are
required to reconcile various experimentally determined
superconducting parameters.

One of the authors (E.E.M.C.) acknowledges R. Prozorov and Q. Chen
for useful discussions. This work was supported by the NSF through
Grant No. DMR9972087. Work at Los Alamos was performed under the
auspices of the U.S. Department of Energy.

\bibliography{CeCoIn5}
\bigskip

\end{document}